\documentclass[10pt,twocolumn,letterpaper]{article}

\usepackage{cvpr}              % To produce the CAMERA-READY version
\usepackage{multirow}
\usepackage[accsupp]{axessibility}
\usepackage{algorithm}      % For algorithm environments
\usepackage{algpseudocode}

\definecolor{cvprblue}{rgb}{0.21,0.49,0.74}
\usepackage[pagebackref,breaklinks,colorlinks,allcolors=cvprblue]{hyperref}

\def\paperID{FedVision-20} % *** Enter the Paper ID here
\def\confName{CVPR}
\def\confYear{2025}

\title{Sporadic Federated Learning Approach in Quantum Environment to Tackle Quantum Noise}
\def\spaces{~~~~~~}
\author{Ratun Rahman \textsuperscript{1} \spaces{} Atit Pokharel \textsuperscript{1} \spaces{} Dinh C. Nguyen  \textsuperscript{1}\\
\textsuperscript{1}Department of Electrical and Computer Engineering, The University of Alabama in Huntsville, USA\\
{\tt\small rr0110@uah.edu, ap1284@uah.edu, dinh.nguyen@uah.edu}
}

\begin{document}
\maketitle
\begin{abstract}
Quantum Federated Learning (QFL) is an emerging paradigm that combines quantum computing and federated learning (FL) to enable decentralized model training while maintaining data privacy over quantum networks.  However, quantum noise remains a significant barrier in QFL, since modern quantum devices experience heterogeneous noise levels due to variances in hardware quality and sensitivity to quantum decoherence, resulting in inadequate training performance.
To address this issue, we propose SpoQFL, a novel QFL framework that leverages sporadic learning to mitigate quantum noise heterogeneity in distributed quantum systems. SpoQFL dynamically adjusts training strategies based on noise fluctuations, enhancing model robustness, convergence stability, and overall learning efficiency. Extensive experiments on real-world datasets demonstrate that SpoQFL significantly outperforms conventional QFL approaches, achieving superior training performance and more stable convergence.
\end{abstract}

\section{Introduction}
\label{sec:intro}
Federated Learning (FL) has emerged as an innovative machine learning framework that enables collaborative model training across decentralized devices while ensuring data privacy \cite{mehta2024securing, ahmed2024fedopt, li2020federated, uddin2024false}.  Traditional centralized machine learning approaches \cite{pokhece} require combining data from several sources on a single server and training the model on the centralized data \cite{rahman2025electrical}.  However, this approach presents two practical challenges: i) \textit{data security} where centralized training exposes sensitive data to security risks since raw data must be sent and kept in a single repository, making it open to breaches and unauthorized access \cite{rahman2024improved,rahman2024multimodal}; and ii) \textit{communication overhead} where sending large and complex datasets to a central server has significant communication costs, particularly in distributed environments with limited bandwidth \cite{bhanbhro2024issues,rahman2024electrical, perf2024}. FL overcomes these limitations by keeping data local on edge devices and sharing model changes only with a central server.  This decentralized strategy improves data security by keeping raw data private to local devices, reducing the risk of data-based attacks.  Furthermore, FL considerably reduces communication costs by sharing only model parameters or gradients rather than complete datasets \cite{karimireddy2020scaffold,mcmahan2017communication}.

Quantum Federated Learning (QFL) combines federated learning (FL) and quantum physics to capitalize on quantum computing's advantages in distributed learning \cite{chehimi2022quantum,park2025entanglement}. Quantum systems excel in handling high-dimensional data \cite{khadka2024microcontroller}, solving complicated optimization problems, and accelerating computations via superposition and entanglement, potentially outperforming conventional approaches \cite{schuld2019quantum,farooq2024enhanced}.  In federated environments, QFL improves data security by keeping sensitive quantum data private in local devices, complying with the quantum no-cloning theorem, which prevents the duplication of quantum states. Furthermore, quantum algorithms enhance optimization and pattern recognition, making QFL ideal for safe, large-scale distributed learning across a wide range of applications \cite{huang2021power}. QFL enables more efficient and scalable decentralized learning by combining the privacy features of FL and the computational power of quantum computing \cite{park2025entanglement,hallaji2024decentralized}.

Despite such promising results, \textit{quantum noise remains a significant challenge for QFL devices}, commonly referred to as noisy intermediate-scale quantum (NISQ) \cite{chen2024robust}. Quantum systems are inherently unstable and susceptible to noise due to hardware errors, environmental interactions, and quantum decoherence, in which qubits lose their quantum states over time \cite{sharma2022implementation}. In quantum computing, noise is generated primarily via gate noise and measurement noise.  Gate noise arises during quantum processes when quantum gates contribute errors due to hardware errors, control inadequacies, or environmental disturbances, resulting in incorrect calculations.  Measurement noise, on the other hand, occurs while extracting classical information from qubits, since quantum measurements are naturally uncertain and error-prone, lowering the reliability of the outcomes \cite{jose2022error}.  These noise effects hinder QFL model training by introducing computational errors, changing learning dynamics, and slowing convergence, making it challenging to maintain training consistency among dispersed quantum devices. In federated environments, where many quantum nodes contribute to training, noise heterogeneity across devices further amplifies learning instability.  Effective noise mitigation is essential for improving the reliability of QFL, maintaining constant training performance, enhancing convergence, and fully exploiting quantum advantages in decentralized learning \cite{wang2022quantumnat,cai2023quantum}.

To address this limitation, in this paper, we propose a novel sporadic QFL approach named \textit{SpoQFL} that effectively reduces the effect of quantum noise in distributed quantum systems. The main contributions of this paper are summarized as follows.

\begin{itemize}
    \item We propose \textit{SpoQFL}, a novel approach for QFL that uses sporadic learning, a strategy for dynamically adapting training techniques to mitigate the effects of quantum noise across distributed quantum devices. 
    \item We consider adaptive and practical quantum noise scenarios and mitigate the noise by selectively altering model updates in response to noise fluctuations
    \item We evaluate the advantage of \textit{SpoQFL} through extensive studies and show that \textit{SpoQFL} surpasses traditional QFL approaches reaching up to 4.87\% higher accuracy on CIFAR-10 and 3.66\% on CIFAR-100, while reducing loss by 16.84\% and 4.15\%, respectively.
\end{itemize}

\section{Related Works}
% \label{sec:formatting}
\subsection{Federated Learning}
FL has become a pivotal approach to distributed machine-learning applications. This framework is particularly valuable in applications where data cannot be centralized due to privacy concerns or communication overheads \cite{rahman2024improved}. Several FL algorithms, such as FedAvg\cite{mehta2024securing} for aggregation, MOON \cite{li2021model} address representation disparity across clients, FedProx \cite{yuan2022convergence} for handling data heterogeneity by introducing a proximal term to stabilize training, FedOpt \cite{ahmed2024fedopt} for enhances convergence through adaptive learning rate rates, and SCAFFOLD \cite{karimireddy2020scaffold} for variance reduction by mitigating client drift, have been developed to enhance FL performance. These methods enable decentralized model training over distributed edge devices, protect data privacy by reducing the need for centralized data aggregation, and improve scalability and security in collaborative learning environments. Despite these advantages, FL algorithms still face challenges due to limited model capacity, slow convergence, and complex data relationships \cite{bhanbhro2024issues}. 

\begin{figure}
    \centering
    \includegraphics[width=0.99\linewidth]{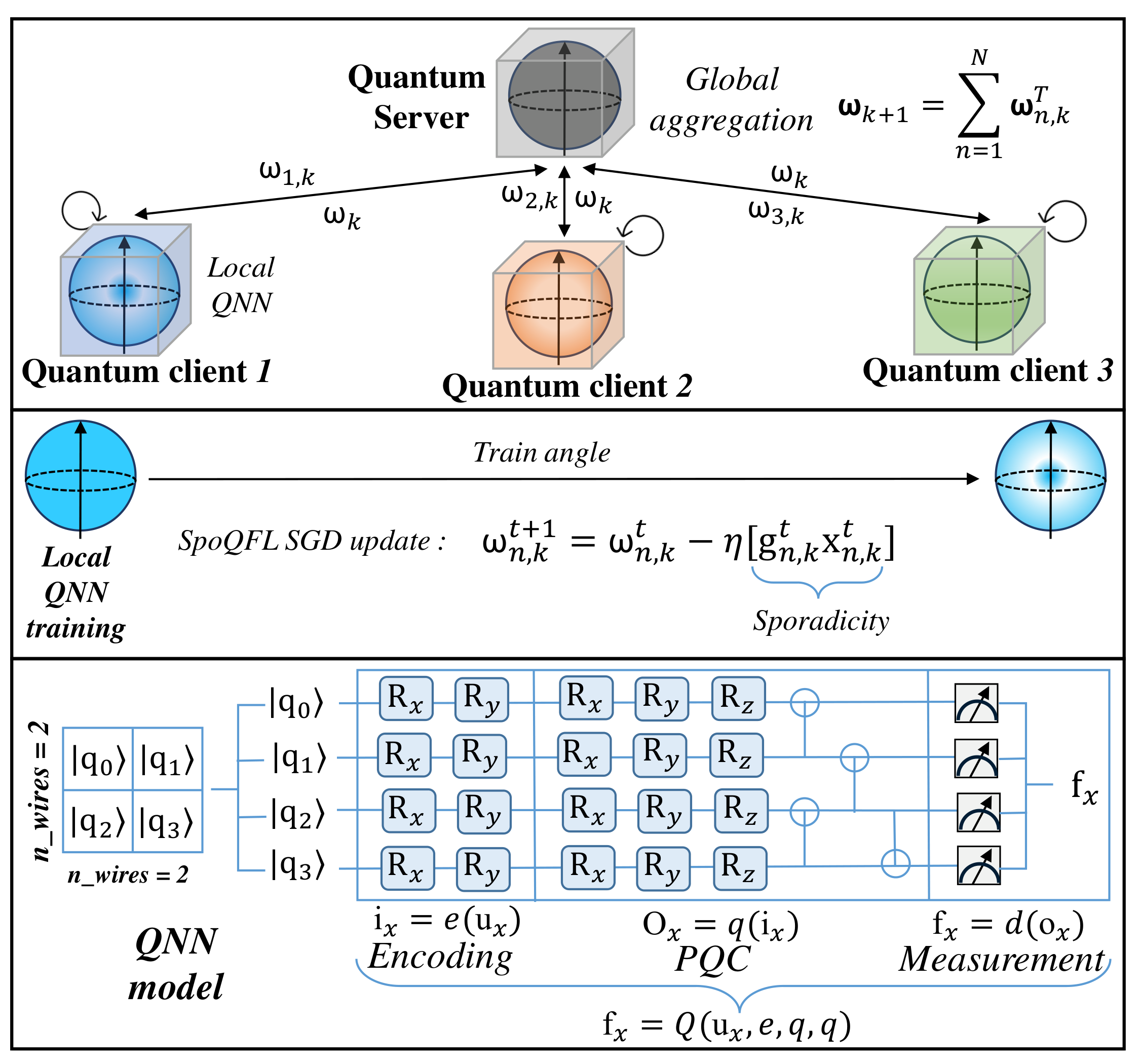}
    \caption{Proposed \textit{SpoQFL} architecture where a set of distributed quantum devices collaborate with a quantum server to train a shared QML model. The proposed framework encompasses sporadic learning that captures noise heterogeneity and mitigates the effects of quantum noise in QFL. }
    \label{fig: overview}
\end{figure}

\subsection{QFL with Quantum Machine Learning}
Quantum machine learning (QML) combines quantum computing with classical machine learning to improve performance in complex computational tasks \cite{biamonte2017quantum}. Similar to the classical FL, QFL allows numerous clients to train models without exchanging raw data, ensuring data privacy and lowering communication overhead \cite{mcmahan2017communication}. By exploiting quantum principles such as qubit's existence in superposition state and entanglements \cite{parmar2023quantum}, multiple research works have shown that QML techniques like quantum neural networks (QNNs) \cite{schuld2019quantum} and quantum support vector machines (QSVMs) \cite{farooq2024enhanced} offer potential speedups and enhanced data representation \cite{huang2021power}. Building on these advancements, QFL extends QML principles to a decentralized learning environment, \cite{park2025entanglement}, while also preserving the FL's inherent data privacy benefits along with improved scalability and performance.  
 QNNs are susceptible to quantum noise caused by decoherence, gate errors, and measurement uncertainty, which distort quantum states and affect model accuracy. There have been extensive research works in mitigating quantum noise effects within the quantum processor itself \cite{wang2022quantumnat,cai2023quantum}. Despite mitigation techniques, residual noise persists specifically in QFL, accumulating as noisy updates from multiple clients propagate during global aggregation which amplifies instability and delays convergence \cite{chen2024robust, dutta2024mqfl}.
 
 Sporadic learning has been explored in classical FL to selectively suppress unstable updates and improve convergence in heterogeneous environments \cite{zehtabi2024decentralized}. However, in the QFL setting, the instability due to quantum noise accumulation and noisy updates is not addressed in prior studies. In this research, we introduce a novel concept of handling such issues with sporadic learning in QFL.
 
%\section{Problem Statement}

\section{Sporadic Quantum Federated Learning Framework}
\textbf{QNN model.}
The quantum model consists of multiple layers containing both single-qubit and multi-qubit gates.  It consists of rotating gates \(R_x, R_y, \text{and} R_z\) and entangling gates (CNOT) that manipulate and entangle qubits to capture complicated data relationships.  These gates are often parameterized and behave similarly to trainable weights in traditional neural networks.  These parameters are adjusted according to a specified loss function.  In the final stage, qubit measurements convert their quantum states into classical bit values, which are then interpreted as probabilities or expectation values.

\subsection{Local Model Training} \label{subsection:local model training}

A QNN model training is initiated by first transforming classical data into quantum states, commonly referred to as quantum encoding. Input data \( w \in \mathbb{R}^{d} \) is first encoded into a quantum state using an encoding unitary \( U_{\text{enc}}(w) \), such that the encoded state $|\psi_{\text{enc}}(w)\rangle$ becomes =$U_{\text{enc}}(w)|0\rangle^{\otimes D}$, where \( D \) represents the number of qubits.

The encoded quantum state is then processed by a parameterized quantum circuit (PQC) composed of multiple layers containing single-qubit rotational gates and two-qubit entanglement gates. The state evolution of the qubits can be expressed in terms of unitary transformations given by
\begin{equation} \label{eq:PQC unitary transformation}
|\psi_{\text{out}}(w, \omega)\rangle = U(\omega)U_{\text{enc}}(w)|0\rangle^{\otimes D}.
\end{equation}

Here, \( U(\omega) = \prod_{d=1}^{L} U_d(\omega_d) \) is the parameterized quantum circuit (PQC) composed of \( L \) layers of parameterized gates.

To evaluate model performance, the output state is measured using a Hermitian observable \( O \), giving the expected value $ f(w, \omega) $ as $\langle\psi_{\text{out}}(w,\omega)|O|\psi_{\text{out}}(w,\omega)\rangle $. In practice, this value is estimated through \( M \) measurement shots, giving the empirical output estimate $\hat{f}_{n,k}(\omega)$ as $\frac{1}{M} \sum_{j=1}^{M} H_j$, where \( H_j \) denotes the individual measurement outcomes.
Then, the loss function $L_{n,k}(\omega)$ is given by $ \ell(y, \hat{f}_{n,k}(\omega))$, where $\ell$ is the categorical cross-entropy loss with $y$ being the respective label of the input data point.
For gradient estimation, the parameter-shift rule is applied:

\begin{equation}\label{eq:parameter shift gradients}
[\hat{g}_{n,k}^{t}]_{d} = \frac{1}{2} \left( \langle\hat{H}\rangle_{|\psi(\omega+\frac{\pi}{2} e_{d})\rangle} - \langle\hat{H}\rangle_{|\psi(\omega-\frac{\pi}{2} e_{d})\rangle} \right),
\end{equation}
where \( e_d \) is the unit vector in the \(d\)-th parameter direction.

The model parameters are updated iteratively using gradient descent:

\begin{equation}\label{eq:gradient update}
\omega_{n,k}^{t+1} = \omega_{n,k}^{t} - \eta \hat{g}_{n,k}^{t},
\end{equation}
where \( \eta \) is the learning rate. After \( T \) local epochs, the final optimized parameters \( \omega_{n,k}^{T} \) are obtained.

\subsection{Federated Learning} \label{subsection:QFL}
FL allows multiple clients to collaboratively train a shared global model without explicitly sharing raw data, ultimately preserving data privacy. In FL, the participating clients share only updated parameters trained in their private local data with a central aggregating server. The server combines these model updates to generate the global model that captures diverse information from multiple trained models. This refined global model is redistributed to clients until a desired performance threshold is met.

QFL extends FL by utilizing QNNs as client models.  In QFL, each quantum client trains a QNN parameterized by 
\(\omega_{n,k}\) that evolves a quantum state 
\(|\psi_{\text{out}}(w, \omega_{n,k})\rangle\) 
as described in \ref{subsection:local model training}.

After performing \( T \) local epochs, each client sends its updated QNN parameters 
\(\omega_{n,k}^T\) to the central server. The server aggregates these updates as 

\begin{equation}\label{eq:global aggregation}
\omega_{k+1} = \frac{1}{N} \sum_{n=1}^{N} \omega_{n,k}^T,
\end{equation}
where \( N \) is the number of participating quantum devices.

QFL provides numerous significant improvements over classical FL. QNNs take advantage of the high-dimensional Hilbert space, resulting in higher expressiveness and faster convergence in specific applications. Furthermore, QFL is particularly useful in applications containing complicated data that classical models may not be able to handle efficiently. QFL, like standard FL, protects data privacy by storing quantum data locally on each device, lowering security threats in distributed quantum computing systems. 

\subsection{Quantum Noise Effect}\label{subsection:noise effect}
Although QFL offers numerous advantages as discussed in \ref{subsection:QFL}, quantum noise significantly affects its performance. Quantum devices, particularly NISQ systems, are highly susceptible to quantum noises from phenomena like decoherence, gate errors, and environmental interference. These noise sources distort quantum states during QNN training, posing challenges to model convergence and stability.

In the \textit{ideal noise-free case}, PQC transforms the initial ground state \( |0\rangle \) into the state \( |\Psi(\omega)\rangle = U(\omega) |0\rangle \), where \(U(\omega)\) is a sequence of parameterized gates defined as
$
\prod_{d=1}^{D} U_d(w_d)V_d.
$
Here, \( U_d(w_d) = \exp \left(-i \frac{w_d}{2} G_d \right) \) represents a parameterized gate, where \( G_d \) is a Pauli string generator (e.g., \( I, X, Y, Z \)) that controls the gate’s effect. The term \( V_d \) is a fixed unitary transformation independent of the trainable parameters \( w_d \).
The PQC of QNN no longer produces a pure state. Instead, the system evolves into a \textit{noisy quantum state}, described by a density matrix:

\begin{equation}\label{eq:density matrix with noise}
\rho_E (\omega) = \tilde{U}_\omega \rho_0 \tilde{U}_\omega^\dagger,
\end{equation}
where \( \rho_0 = |0\rangle \langle 0| \) is the initial pure state, and \( \tilde{U}_\omega \) is the noise-affected PQC transformation. The symbol \( \tilde{U}_\omega^\dagger \) represents the Hermitian adjoint (complex conjugate transpose) of \( \tilde{U}_\omega \), ensuring the resulting density matrix remains Hermitian and trace-preserving.

A widely used noise model for practical systems is the \textit{Pauli noise channel}, which introduces probabilistic disturbances $E(\rho)$ during gate operations given by $(1 - \epsilon)\rho + \epsilon \sum_j E_j \rho E_j^\dagger$.
% \begin{equation}
% E(\rho) = (1 - \epsilon)\rho + \epsilon \sum_j E_j \rho E_j^\dagger
% \end{equation}
Here, \( \epsilon \) is the probability of error, and \( E_j \) are Pauli operators that introduce noise during gate execution.
Quantum noise also distorts the empirical loss function. Since QNN training relies on measurement outcomes, the noisy loss function $\hat{L}(\omega)$ is estimated by averaging \(M\) measurement shots as $\frac{1}{M} \sum_{j=1}^{M} H_j$.
% \begin{equation}
% \hat{L}(\omega) = $\frac{1}{M} \sum_{j=1}^{M} H_j$
% \end{equation}
 Here, \( H_j \) is the outcome of the \(j\)-th shot for an observable \( H \).

Despite averaging multiple shots, residual noise remains and propagates into gradient estimates. Under noisy conditions, this estimated gradient as in \eqref{eq:parameter shift gradients} becomes:

\begin{equation}\label{eq:gradient with noise}
\hat{g}_{n,k}^t = \nabla f(\omega_{n,k}) + \xi_{n,k}^t.
\end{equation}

Here, \( \nabla f(\omega_{n,k}) \) is the noise-free gradient, and \( \xi_{n,k}^t \) is a noise-induced deviation. This deviation follows a bounded variance given by

\begin{equation}\label{eq:variance of noise}
\text{Var}(\xi_{n,k}^t) \leq \frac{\nu N_h D \text{Tr}(H^2)}{2M},
\end{equation}
wshere \( \nu \) is a noise-related constant, \( N_h \) denotes the number of Hermitian observable terms, \( D \) is the number of qubits, and \( M \) represents the number of measurement shots. Noise accumulates over multiple rounds, destabilizing convergence. After \( T \) communication rounds, the noise-induced error bound can be formulated as

\begin{equation}\label{eq:error bound of noise in loss}
\xi_t = E[L(\omega_T)] - L^* \leq (1 - \eta \mu)^T (E[L(\omega_0)] - L^*) + \frac{1}{2} \frac{\eta L V}{\mu},
\end{equation}
where \( L^* \) is the optimal loss, \( \eta \) is the learning rate, \( \mu \) is the strong convexity constant, and \( V \) is the noise variance bound. This result highlights the cumulative impact of noise on convergence, as both measurement uncertainty and gate-induced noise gradually degrade QFL performance.

\subsection{Quantum Noise Mitigation with Sporadic Learning}
We anticipate that the major challenge in QML model training within the NISQ-based QFL system is the presence of quantum noise which is accumulated in multiple communication rounds of QFL during training, \ref{subsection:noise effect}, degrading the global model's stability.
\textit{Sporadic learning} mitigates this effect by dynamically scaling updates based on observed noise intensity. In QFL, each client’s gradient estimate is inherently susceptible to noise as formulated in \eqref{eq:gradient with noise}. To minimize the impact of unstable updates, sporadic learning introduces a scaling factor, termed the \textit{sporadic variable}, that dynamically attenuates noise-heavy updates. Given the noise deviation \( \xi_{n,k}^t \) in a client’s gradient estimate, the sporadic variable is defined as

\begin{equation}\label{eq:sporadic variable}
x_{n,k}^t = \exp\bigl(-\gamma\,\lvert \xi_{n,k}^t \rvert\bigr),
\end{equation}
where \( \gamma \) is a tunable parameter that determines the suppression strength. For clients experiencing high noise, \( x_{n,k}^t \) becomes small, limiting the impact of unstable updates. Conversely, for minimal noise conditions, \( x_{n,k}^t \approx 1 \), ensuring stable updates proceed normally.
 
Each client’s updated parameters are scaled accordingly before being sent to the server, expressed as:

\begin{equation}\label{eq:gradient update with sporadic scaling}
\omega_{n,k}^{t+1} = \omega_{n,k}^{t} - \eta \bigl( \hat{g}_{n,k}^t \cdot x_{n,k}^t \bigr).
\end{equation}

This adaptive adjustment ensures that devices with excessive noise contribute less to the global model. As described in ~\ref{subsection:QFL}, the server aggregates these stabilized local updates using the standard FL aggregation rule in \eqref{eq:global aggregation}.

Since sporadic learning directly reduces the noise variance term \( V \) in the previously established error bound in \eqref{eq:error bound of noise in loss}, it improves both convergence stability and model performance in practical QFL environments.

\subsection{Proposed \textit{SpoQFL} Algorithm}
The proposed \textit{SpoQFL} algorithm in Algorithm \ref{algo:spoQFL} begins by initializing the global model and local parameters (lines 2-6). Then the clients get trained on local data with noisy gradient estimation (line 8) and compute the sporadic variable based on noise intensity (line 9). Subsequently, updates are skipped if noise exceeds a defined threshold (lines 10-11), while stable updates are scaled to limit noise propagation (line 13). The stabilized models are sent to the server (lines 16-19), which aggregates, evaluates, and broadcasts the updated global model to all the clients. This process repeats for a predefined number of global rounds specified in line 1, achieving the optimal global model. An overview of the \textit{SpoQFL} framework is shown is figure \ref{fig: overview}.
\begin{algorithm}[!ht]
\footnotesize
    \caption{Proposed \textit{SpoQFL} Algorithm}
    \label{algo:spoQFL}
    \begin{algorithmic}[1]
        \State \textbf{Input:} Global communication rounds \( \mathcal{K} \), local training epochs \( \mathcal{T} \), set of quantum devices \( \mathcal{N} \), learning rate \( \eta \), sporadic control parameter \( \gamma \), noise threshold \( \tau \)
        \State \textbf{Initialization:} Initialize global model parameters \( \boldsymbol{\omega}_0 \)
        \For{each global communication round \( k \in \mathcal{K} \)}
            \State Distribute current global model \( \boldsymbol{\omega}_{k} \) to all devices
            \For{each device \( n \in \mathcal{N} \) in parallel}
                \State Set initial local model parameters \( \boldsymbol{\omega}_{n,k}^0 = \boldsymbol{\omega}_{k} \)
                \For{each local training epoch \( t \in \mathcal{T} \)}
                    \State Compute gradient estimate:
                    \[
                    \hat{g}_{n,k}^t = \nabla f_{n} (\boldsymbol{\omega}_{n,k}^t) + \xi_{n,k}^t
                    \]
                    \State Compute sporadic variable:
                    $
                    x_{n,k}^t$ using \eqref{eq:sporadic variable}.
                    \If{\( x_{n,k}^t < \tau \)} 
                        \State \textbf{Skip update} (Suppresses unstable updates)
                    \Else
                        \State Calculate loss and update local model parameters:
                        \[
                        \boldsymbol{\omega}_{n,k}^{t+1} = \boldsymbol{\omega}_{n,k}^{t} - \eta \bigl( \hat{g}_{n,k}^t \cdot x_{n,k}^t \bigr)
                        \]
                    \EndIf
                \EndFor
                \State Transmit updated model parameters \( \boldsymbol{\omega}_{n,k}^T \) to the server
            \EndFor
            \State Server aggregates client updates using \eqref{eq:global aggregation} to get \(\boldsymbol{\omega}_{k+1} \).
            % \State \textbf{Convergence Check:} Stop if \( | L(\boldsymbol{\omega}_{k+1}) - L(\boldsymbol{\omega}_{k}) | \leq \epsilon \)
            \State Server evaluates the global model performance and broadcasts \( \boldsymbol{\omega}_{k+1} \) to all clients
        \EndFor
        \State \textbf{Output:} Optimal global model \( \boldsymbol{\omega}^* \) after \( K \) rounds
    \end{algorithmic}
\end{algorithm}

\begin{figure*}[ht!]
    \centering
    \footnotesize
    \begin{subfigure}[t]{0.245\linewidth} % Adjust the width to occupy half of the column width
        \centering
        \includegraphics[width=\linewidth]{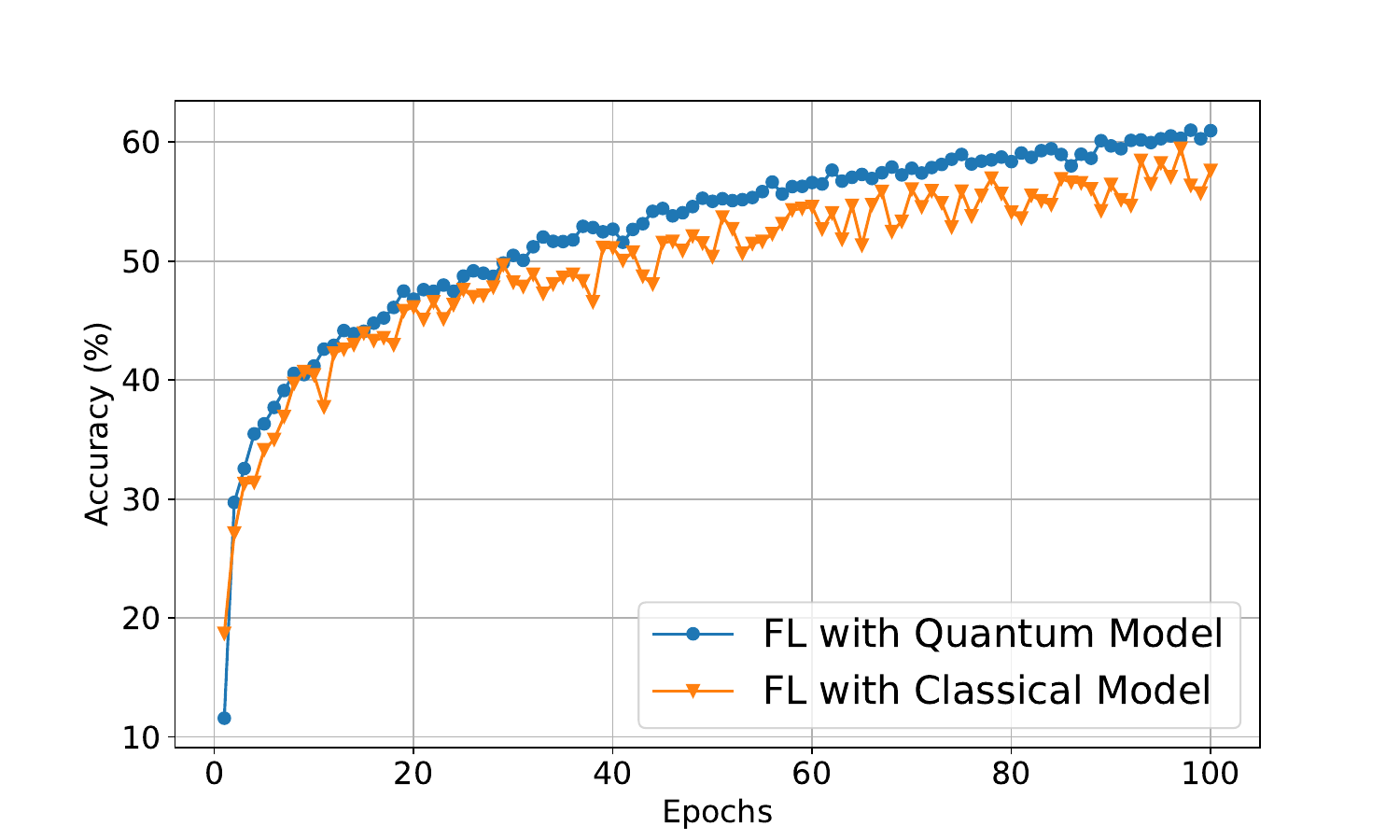}
        \caption{\footnotesize CIFAR-10 IID accuracy.}
    \end{subfigure}
    \hfill % Add horizontal space between subfigures
    \begin{subfigure}[t]{0.245\linewidth} % Adjust the width to occupy half of the column width
        \centering
        \includegraphics[width=\linewidth]{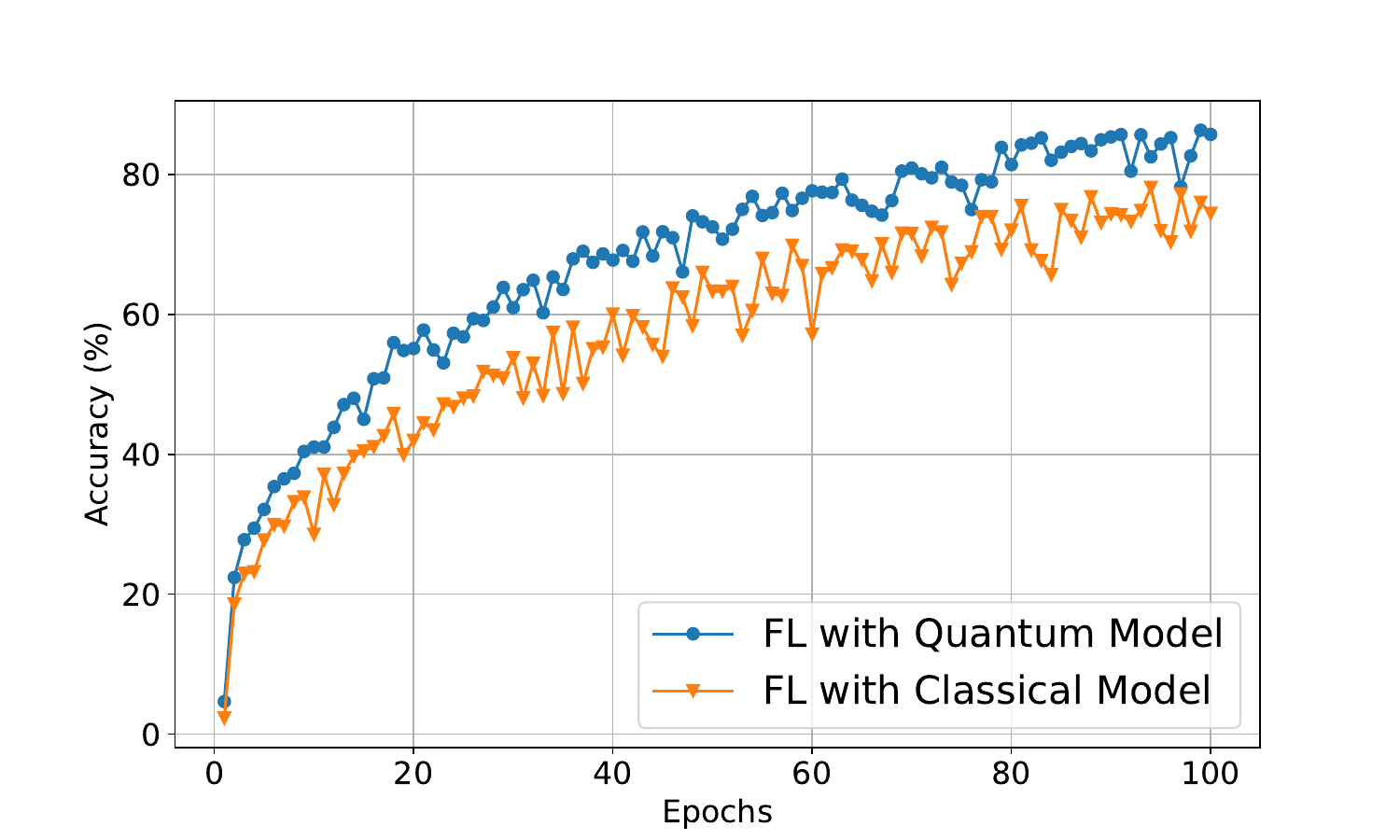}
        \caption{\footnotesize CIFAR-10 Non-IID accuracy.}
    \end{subfigure}
    \hfill % Add horizontal space between subfigures
    \begin{subfigure}[t]{0.245\linewidth} % Adjust the width to occupy half of the column width
        \centering
        \includegraphics[width=\linewidth]{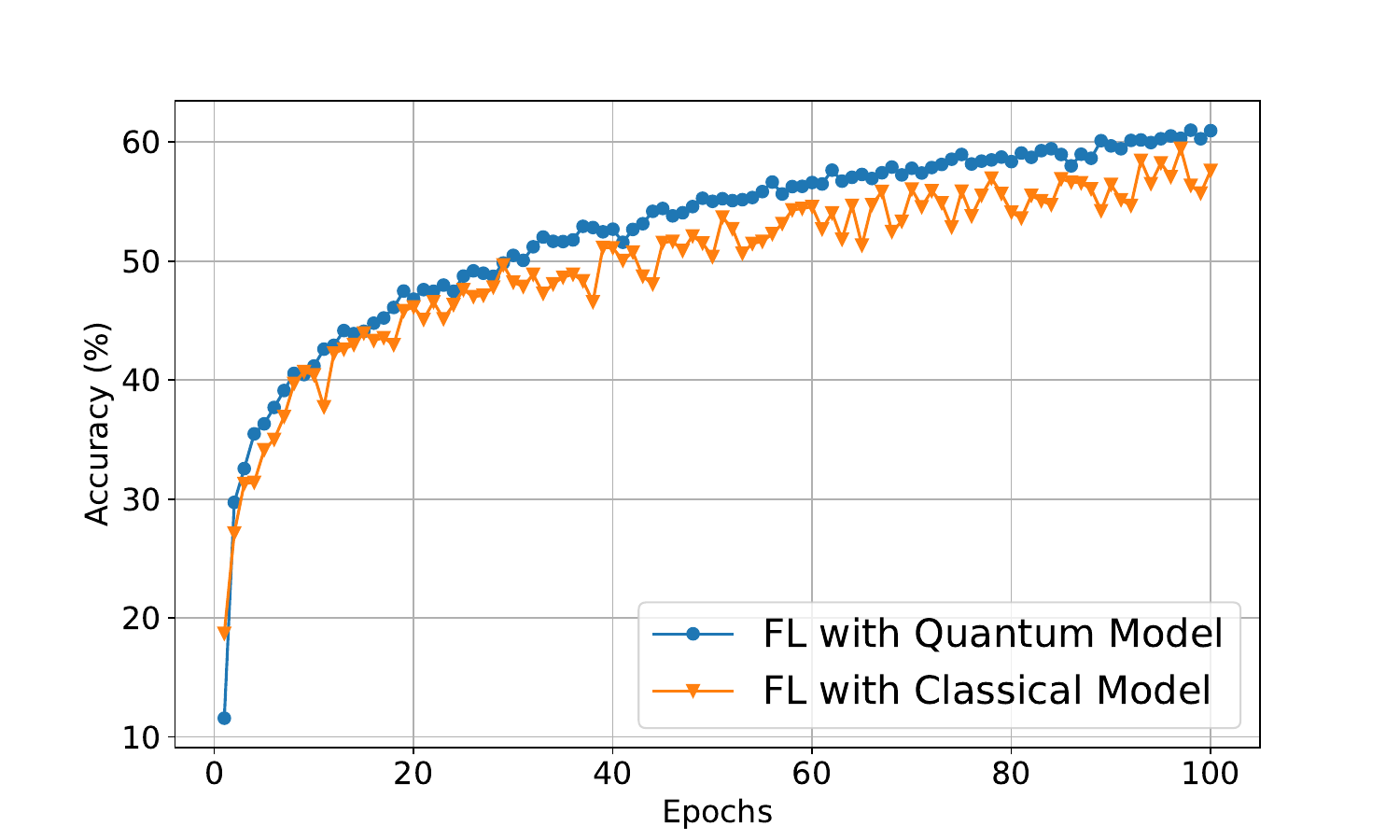}
        \caption{\footnotesize CIFAR-100 IID accuracy.}
    \end{subfigure}
    \hfill
    \begin{subfigure}[t]{0.245\linewidth} % Adjust the width to occupy half of the column width
        \centering
        \includegraphics[width=\linewidth]{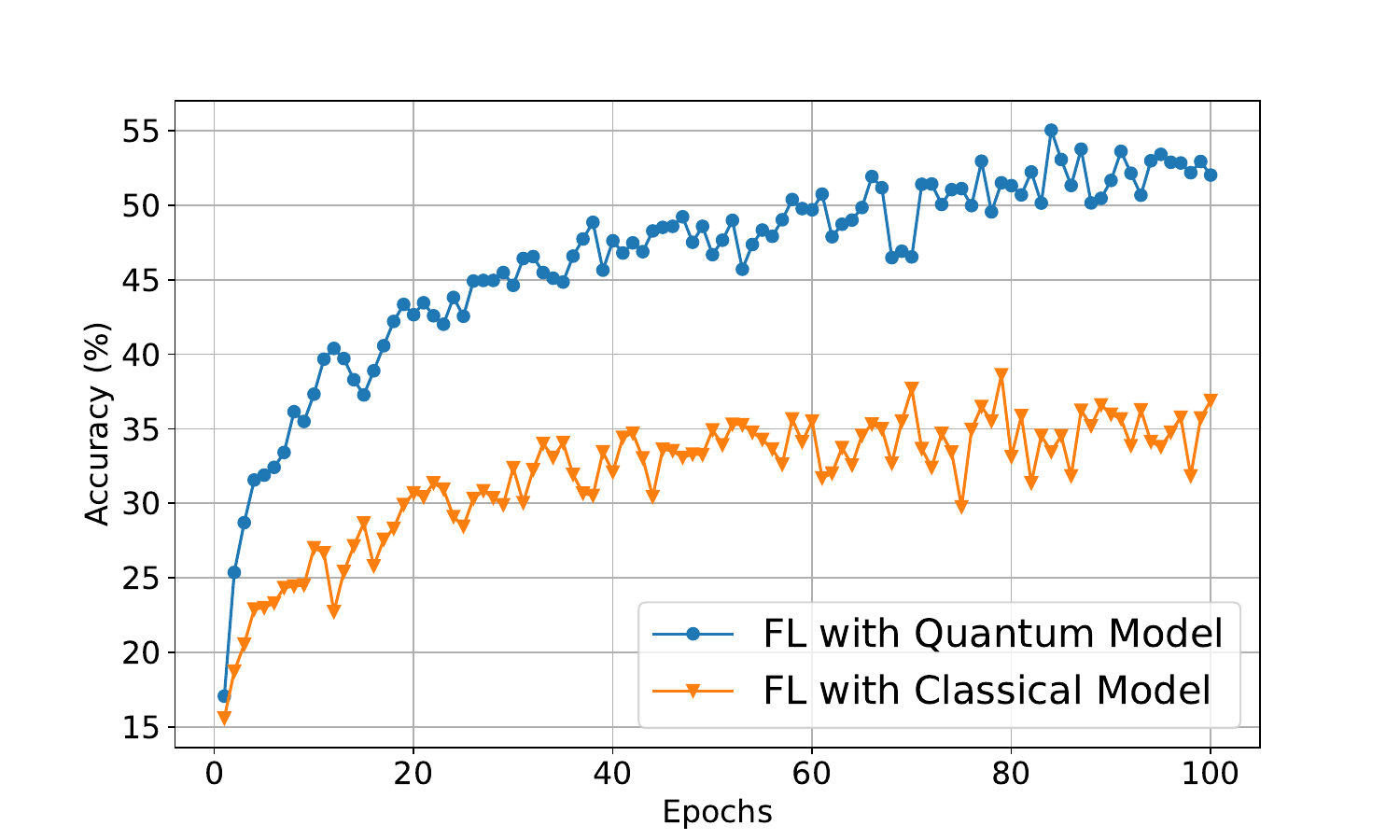}
        \caption{\footnotesize CIFAR-10 Non-IID accuracy.}
    \end{subfigure}
    \caption{Performance comparison between FL with classical model and FL with the quantum model for CIFAR-10 and CIFAR-100 dataset in IID and non-IID data distribution. The simulation results demonstrate the effects of using the quantum model.}
    \label{fig: 10comp}
\end{figure*}

\section{Experiments}
\subsection{Dataset}
To evaluate the performance of \textit{SpoQFL}, we have used two baseline datasets: CIFAR-10 and CIFAR-100 \cite{krizhevsky2009learning}. The CIFAR-10 dataset contains 60,000 color images (32×32 pixels) divided evenly over 10 classes: airplane, car, bird, cat, deer, dog, frog, horse, ship, and truck.  It comprises 50,000 training and 10,000 test images, with 6,000 examples in each class.  The CIFAR-100 dataset has a similar structure, but it contains 100 classes divided into 20 superclasses, each with 600 pictures (500 for training and 100 for testing).  CIFAR-100 presents a more fine-grained classification problem due to its larger number of categories.

We initially normalize the pixel values to the range [0,1] to increase numerical stability and training efficiency.  To emulate real-world federated learning settings, we partition the datasets into non-IID subsets using label-skewed and Dirichlet-based partitioning \cite{hsu2019measuring}, guaranteeing that each client receives a distinctive yet imbalanced set of classes. Standard data augmentation approaches, such as random cropping, horizontal flipping, and random rotation, are used to improve model generalization.  Furthermore, to improve computational performance, we introduce batch-wise data loading and ensure a balanced training procedure among FL clients.

\subsection{Simulation Settings}
Our \textit{SpoQFL} framework consists of a centralized quantum server and ten NISQ devices.  Each quantum device is in charge of training a QNN for image classification.  The framework uses quantum state preparation techniques to encode input data, such as pixel values, into quantum states via methods such as amplitude encoding or angle encoding. For quantum simulations, we use the \textit{torchquantum} library.  The training procedure incorporates conventional optimization strategies to modify quantum gate settings, as well as gradient-based approaches that take into consideration the system's quantum nature.  This configuration uses quantum mechanical features, such as superposition and entanglement, to improve computing efficiency and performance.

\textbf{Non-IID data distribution.} To examine heterogeneous quantum encoding, we simulate a QFL framework in which clients use heterogeneous encoding algorithms, resulting in varying quantum state representations.  We examine \(N=10 \) clients, each using a unique encoding method, including basis, amplitude, phase, angle, and entanglement encoding.  Clients train on distinct encoded quantum states \(\rho_i^{(k)} \), leading in varied feature spaces that affect training stability and model convergence. We assess the state fidelity across client datasets to quantify this heterogeneity, and we find an average pairwise fidelity variation of \( 0.65 \), emphasizing the differences brought by various encoding methods. This heterogeneity further extends into the quantum environment via quantum encoding, in which classical values are directly translated to quantum states, resulting in further variations in the quantum data.

\subsection{Simulation Results on QFL}
\textbf{Effects on using the quantum model.} For our first simulation, we compare two FL approaches: FL with classical model and FL with quantum models for dataset CIFAR-10 and CIFAR-100 in figure \ref{fig: 10comp}. The simulation results clearly demonstrate the superiority of QFL over traditional FL in terms of both performance and convergence in both IID and non-IID data distribution. Furthermore, QFL was able to produce more stable results even in the non-IID data environment. In the consequent simulations, we proceed with FL using the QNN model. 

\textbf{Ablation study.} In our ablation study, we investigate how different quantum factors affect the performance of our QFL model, with a particular emphasis on the number of quantum bits (qubits) and quantum layers.  Table \ref{tab: qubits} summarizes our experiments that compared model performance across multiple configurations of qubits $D_q$ during each communication round.  In those instances, all parameters and training epochs are similar, with the only difference being the number of qubits, which were set to 2, 3, 5, and 10.  The findings show that increasing the amount of qubits improves the model's performance, with $D_q = 10$ producing the best results for both datasets. Further increasing qubits exponentially expands the quantum state, so additional simulations use 10 qubits to balance performance and complexity. 
\begin{table}
    \centering
    \footnotesize
    \begin{tabular}{|c|cc|cc|}
    \toprule
    \multicolumn{1}{|c|}{\multirow{2}{*}{$D_q$ (Num Qubit)}} & \multicolumn{2}{c|}{\textbf{CIFAR-10}} & \multicolumn{2}{c|}{\textbf{CIFAR-100}} \\
    \cmidrule{2-5}
    & \textbf{Loss Value} & \textbf{Acc.} & \textbf{Loss Value} & \textbf{Acc.} \\
    \midrule
    2  & 2.266 & 14.58\%  & 4.026 & 11.54\% \\
    3  & 1.724 & 23.80\%  & 3.523 & 21.45\% \\
    5  & 1.051 & 36.78\%  & 3.011 & 29.78\% \\
    10 & \textbf{0.666} & \textbf{62.98\%} & \textbf{1.320} & \textbf{55.04\%} \\
    \bottomrule
    \end{tabular}
    \caption{Difference in performance in QFL with the different number of qubits $D_q$ across various datasets. The qubits being compared were 2, 3, 5, and 10.  We limited the number of qubits to 10 to minimize excessive computational demands and resource consumption and 1 qubit unable to process the complex image.}
    \label{tab: qubits}
\end{table}

As an extension of our ablation studies, we analyze the effect of varying the number of quantum layers (\(l\)) in our QFL model. Table \ref{tab: qlayer} presents our experiments with layer configurations of 1, 2, 3, 5, and 10. We ensure consistent experimental conditions, including the number of training rounds and the number of qubits (\(D_c = 10\)). Unlike our observations with qubit variations, modifying the number of quantum layers does not exhibit a straightforward linear correlation with model performance. Interestingly, the configuration with \(l = 3\) layer for CIFAR-10 and \(l=1\) layer for CIFAR-100 outperforms the others. Based on these findings, we select \(l = 1\) as the quantum layer configuration for subsequent evaluations to optimize our model's performance.

\begin{table}
    \centering
    \footnotesize
    \begin{tabular}{|c|cc|cc|}
    \toprule
    \multicolumn{1}{|c|}{\multirow{2}{*}{$l$ (num layers)}} & \multicolumn{2}{c|}{\textbf{CIFAR-10}} & \multicolumn{2}{c|}{\textbf{CIFAR-100}} \\
    \cmidrule{2-5}
    & \textbf{Loss Value} & \textbf{Acc.} & \textbf{Loss Value} & \textbf{Acc.} \\
    \midrule
    1  & 0.6747  & 87.75\%  & \textbf{1.3388}  & \textbf{55.63\%}  \\
    2  & 0.6774  & 86.55\%  & 1.3716  & 53.18\%  \\
    3  & \textbf{0.5806}  & \textbf{89.36\%}  & 1.3660  & 52.98\%  \\
    4  & 0.7749  & 85.75\%  & 1.4045  & 52.02\%  \\
    5  & 0.7261  & 85.67\%  & 1.4935  & 48.34\%  \\
    10 & 1.1392  & 72.46\%  & 1.8187  & 36.89\%  \\
    \bottomrule
    \end{tabular}  
    \caption{Difference in performance in QFL with different numbers of layers $l$ across CIFAR-10 and CIFAR-100 datasets.The study investigates the trade-offs between model expressiveness and training efficiency, arguing that fewer layers may result in underfitting due to limited expressiveness, whereas more layers improve representational capability at the expense of encountering barren plateaus that impede efficient optimization.}
    \label{tab: qlayer}
\end{table}
\begin{table}
    \centering
    \footnotesize
    \begin{tabular}{|c|cc|cc|}
    \toprule
    \multirow{2}{*}{Loss function} & \multicolumn{2}{c|}{\textbf{CIFAR-10}} & \multicolumn{2}{c|}{\textbf{CIFAR-100}} \\
    \cmidrule{2-5}
    & \textbf{Train Acc.} & \textbf{Test Acc.} & \textbf{Train Acc.} & \textbf{Test Acc.} \\
    \midrule
    CrossEntropy  & \textbf{86.12\%}  & \textbf{82.93\%}  & \textbf{54.43\%}  & \textbf{52.60\%}  \\
    MSE  & 70.15\% & 65.53\%  & 47.32\% & 44.39\%  \\
    BCE  & 79.97\% & 75.13\% & 54.02\% & 48.13\%  \\
    \bottomrule
    \end{tabular}  
    \caption{Performance comparison between 3 different loss functions: CrossEntropy Loss, MSE Loss, and BCE Loss in selected datasets. We compare multiple loss functions to see which one improves model accuracy the most effectively for certain tasks and data. }
    \label{tab: comp_loss}
\end{table}
\begin{table}
\footnotesize
\centering
\begin{tabular}{|c|p{3cm}|c|c|c|}
\hline
 & \multirow{2}{*}{Learning rate} & \multicolumn{2}{c|}{CrossEntropy Loss} \\
 \cline{3-4}
 & & Loss & Accuracy \\
 \hline
 \multirow{4}{*}{\scriptsize{\rotatebox{90}{CIFAR-10}}}
 & learning rate = 0.5 & \textbf{0.4362} & \textbf{84.82\%} \\
 \cline{2-4}
 & learning rate = 0.1 & 0.705 & 82.32\%  \\
 \cline{2-4}
 & learning rate = 0.01 & 2.204 & 51.19\%\\
 \cline{2-4}
 & learning rate = 0.001 & 3.6568 & 22.42\%  \\
 \hline
 \multirow{4}{*}{\scriptsize{\rotatebox{90}{CIFAR-100}}}
 & learning rate = 0.5 & 2.3211 & 14.80\% \\
 \cline{2-4}
 & learning rate = 0.1 & \textbf{1.3639} & \textbf{53.76\%}  \\
 \cline{2-4}
 & learning rate = 0.01 & 1.6247 & 45.30\%\\
 \cline{2-4}
 & learning rate = 0.001 & 2.005 & 32.09\%  \\
 \hline
\end{tabular}
\caption{Difference in performance using different learning rates. The
learning rates include 0.5, 0.1, 0.01, and 0.001 after 100 epochs for both CIFAR-10 and CIFAR-100. Learning rates substantially influences the efficiency and outcomes of the training process.}
\label{table: comp_lr}
\end{table}
\begin{table*}
    \centering
    \footnotesize
    \begin{tabular}{|c|c|cc|cc|}
    \toprule
    \multirow{2}{*}{\textbf{\shortstack{Data\\distribution}}}&\multirow{2}{*}{\textbf{\shortstack{Number\\of Clients}}} & \multicolumn{2}{c|}{\textbf{CIFAR-10}} & \multicolumn{2}{c|}{\textbf{CIFAR-100}} \\
    \cmidrule{3-6}
    & & \textbf{Loss value} & \textbf{Accuracy} & \textbf{Loss value} & \textbf{Accuracy} \\
    \midrule
    IID & 3  & 0.6721 & 87.92\%  & 1.4123 & 53.25\% \\
    & 5  & 0.6347  & 88.45\% & 1.3789 & 54.89\% \\
    & 10  & \textbf{0.5983}  & \textbf{89.67\%} & \textbf{1.3452}  & \textbf{56.78\%}  \\
    \midrule
    Non-IID & 3  & 0.7342   &85.43\%  & 1.5689  & 50.12\%  \\
    & 5  &0.6985  &  86.21\%  & 1.5023   & 51.76\%  \\
    & 10  & \textbf{0.6654}  & \textbf{ 87.03\%} & \textbf{1.4568  }  & \textbf{52.98\%}  \\
    \bottomrule
    \end{tabular}  
    \caption{Impact of varying the number of clients on QFL performance. The number of clients used for the comparison is 3, 5, and 10 for both IID and non-IID in CIFAR-10 and CIFAR-100 datasets. }
    \label{tab:num_clients}
\end{table*}
\textbf{Effects on different loss functions.}
In our next line of experiments, we examine how different loss functions affect model performance on the CIFAR-10 and CIFAR-100 datasets in Table~\ref{tab: comp_loss}.  CrossEntropy Loss produces the highest accuracies: 86.12\% on CIFAR-10 and 54.43\% on CIFAR-100 during training, and 82.93\% and 52.60\% in testing.  In contrast, MSE and BCE Losses demonstrate decreased effectiveness, particularly with MSE on CIFAR-10, where training and testing accuracies were only 70.15\% and 65.53\%, respectively.  This demonstrates CrossEntropy's strengths and effectiveness in performing classification tasks on these image datasets.

\textbf{Effects on different learning rates.}
Table~\ref{table: comp_lr} shows how altering learning rates affect the performance of a model using CrossEntropy Loss on the CIFAR-10 and CIFAR-100 datasets.  In CIFAR-10, the best performance is obtained with a learning rate of 0.5, which results in a loss of 0.4362 and an accuracy of 84.82\%.  In contrast, as the learning rate declines to 0.1, 0.01, and 0.001, both loss and accuracy, with the best performance occurring at higher learning rates.  For CIFAR-100, the ideal learning rate is 0.1, resulting in the lowest loss of 1.3639 and the maximum accuracy of 53.76\%.  Lower learning rates (0.01 and 0.001) result in more losses and lower accuracies, while the greatest rate (0.5) causes a considerable decline in accuracy to 14.80\%.

\textbf{Effects on the number of clients.} 
Table \ref{tab:num_clients} shows how varying numbers of clients affect QFL performance for CIFAR-10 and CIFAR-100 in IID and Non-IID distributions.  As the number of clients increases from 3 to 10, the loss for CIFAR-10 decreases from 0.6721 to 0.5983, while accuracy climbs from 87.92\% to 89.67\%.  Similarly, for CIFAR-100, the loss falls from 1.4123 to 1.3452, but accuracy rises from 53.25\% to 56.78\%.  The IID configuration regularly outperforms the non-IID condition, demonstrating the effect of data heterogeneity on training.  Despite this, increasing the number of clients improves overall performance and reinforces QFL's scalability, while good aggregation mechanisms are still required to handle non-IID circumstances.

\subsection{Simulation Results on Quantum Noise}
In this line of research, we examine the negative effect of quantum noise and how sporadic learning can solve this problem.

\textbf{Effects on quantum noise.} 
Table \ref{tab:noiselevel} shows how different noise levels (\(\epsilon\)) affect model accuracy for CIFAR-10 and CIFAR-100 datasets under IID and non-IID settings.  The noise levels range from 0.001 to 0.5, with higher values providing additional uncertainty in the training process. As expected, model accuracy decreases as the level of noise increases.  In the IID setup, CIFAR-10 accuracy decreases from 88.93\% at \(\epsilon = 0.001\) to 72.31\% at \(\epsilon = 0.5\), while CIFAR-100 also decreases from 56.70\% to 39.21\%.  The Non-IID scenario has a more dramatic degradation, with CIFAR-10 dropping from 86.87\% to 68.45\% and CIFAR-100 reducing from 52.91\% to 35.67\%.  This emphasizes the difficulty of training models in non-IID situations, where data heterogeneity increases performance loss due to noise.

\begin{table}
\footnotesize
    \centering
    \begin{tabular}{|c|c|c|c|}
    \hline
        \textbf{Data} & \textbf{Noise} & \textbf{CIFAR-10} & \textbf{CIFAR-100}\\
        \textbf{distribution} & \textbf{level ($\epsilon$)} & \textbf{Accuracy} & \textbf{Accuracy}\\
    \hline
        \multirow{4}{*}{IID} & 0.001 & \textbf{88.93\%} & \textbf{56.70\%} \\ \cline{2-4}
        & 0.01  & 86.75\% & 54.32\% \\  \cline{2-4}
        & 0.1   & 81.42\% & 48.76\% \\  \cline{2-4}
        & 0.5   & 72.31\% & 39.21\% \\ \hline
        \multirow{4}{*}{Non-IID} & 0.001 & 86.87\% & 52.91\% \\ \cline{2-4}
        & 0.01  & \textbf{84.12\%} & \textbf{50.10\%} \\  \cline{2-4}
        & 0.1   & 78.89\% & 44.23\% \\  \cline{2-4}
        & 0.5   & 68.45\% & 35.67\% \\ \hline
    \end{tabular}
    \caption{Impact of noise level ($\epsilon$) on CIFAR-10 and CIFAR-100 accuracy under IID and Non-IID data distributions. The noise level considered for comparison is 0.001, 0.01, 0.1, and 0.5.}
    \label{tab:noiselevel}
\end{table}

\begin{figure*}[ht!]
    \centering
    \footnotesize
    \begin{subfigure}[t]{0.245\linewidth} % Adjust the width to occupy half of the column width
        \centering
        \includegraphics[width=\linewidth]{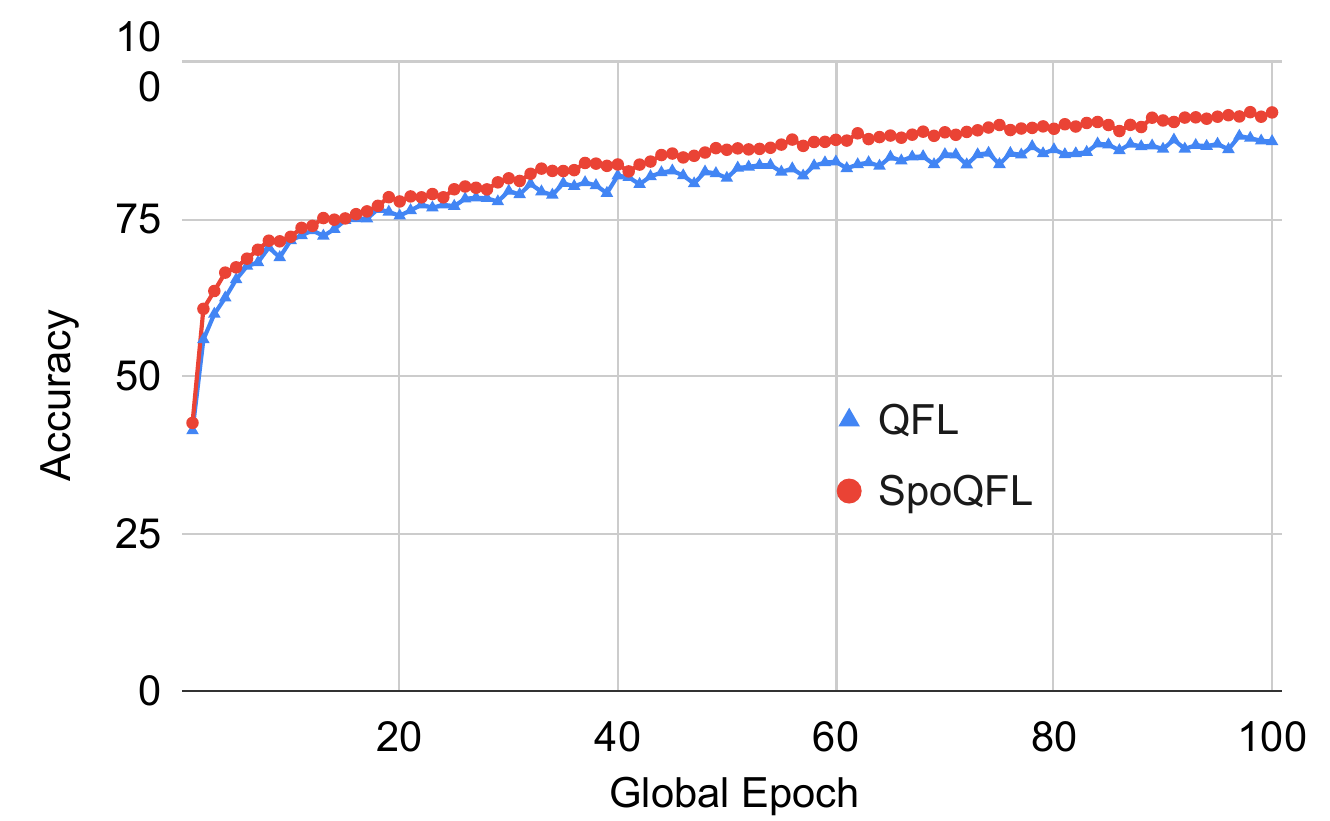}
        \caption{\footnotesize CIFAR-10 IID accuracy.}
    \end{subfigure}
    \hfill % Add horizontal space between subfigures
    \begin{subfigure}[t]{0.245\linewidth} % Adjust the width to occupy half of the column width
        \centering
        \includegraphics[width=\linewidth]{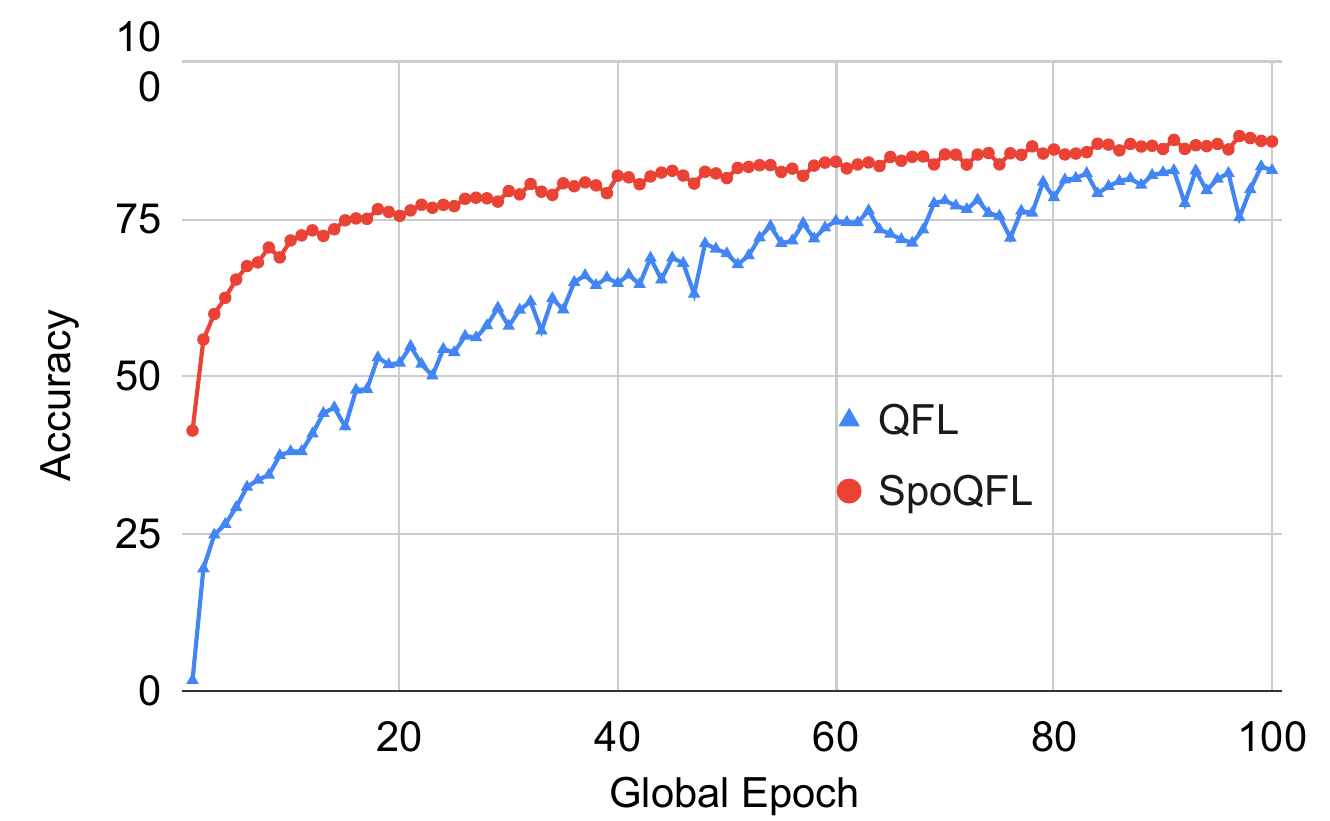}
        \caption{\footnotesize CIFAR-10 Non-IID accuracy.}
    \end{subfigure}
    \hfill % Add horizontal space between subfigures
    \begin{subfigure}[t]{0.245\linewidth} % Adjust the width to occupy half of the column width
        \centering
        \includegraphics[width=\linewidth]{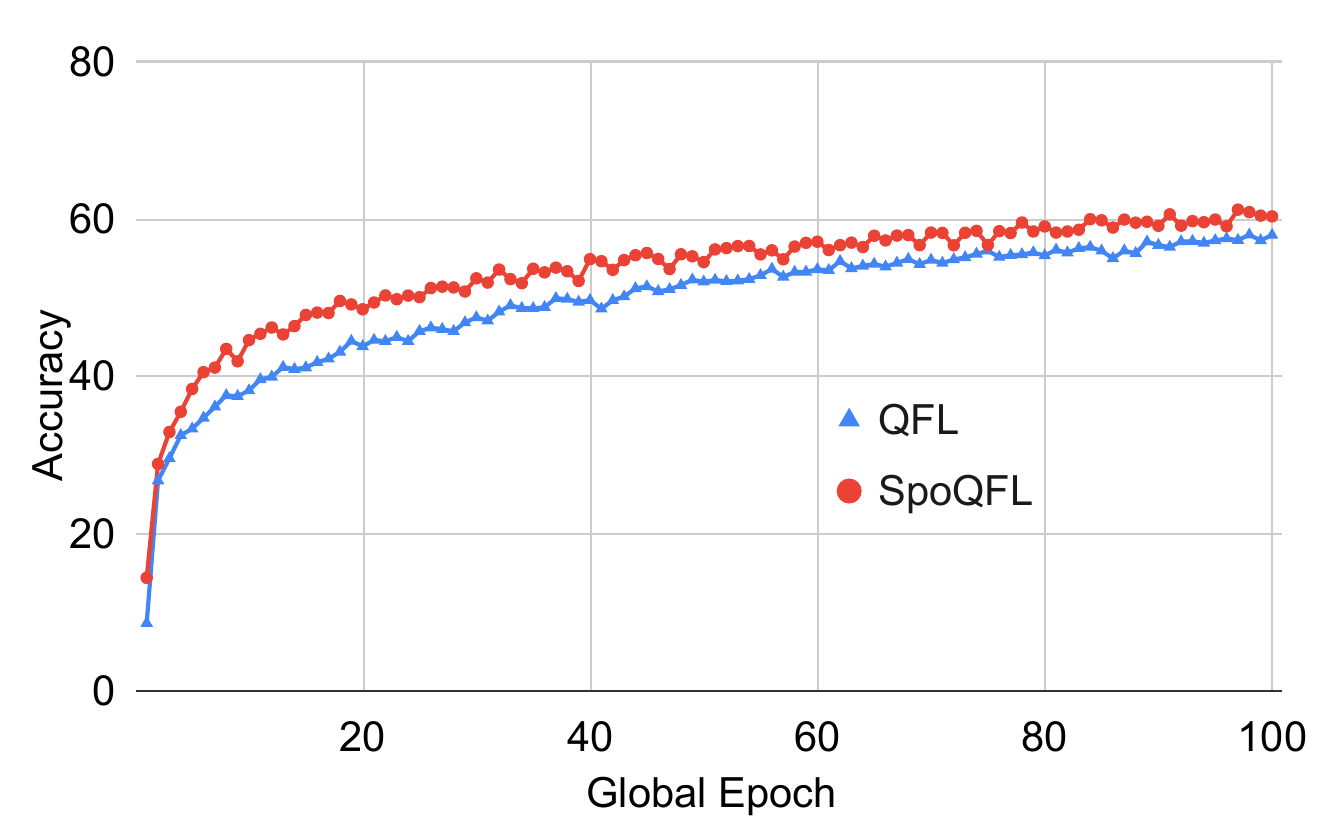}
        \caption{\footnotesize CIFAR-100 IID accuracy.}
    \end{subfigure}
    \hfill
    \begin{subfigure}[t]{0.245\linewidth} % Adjust the width to occupy half of the column width
        \centering
        \includegraphics[width=\linewidth]{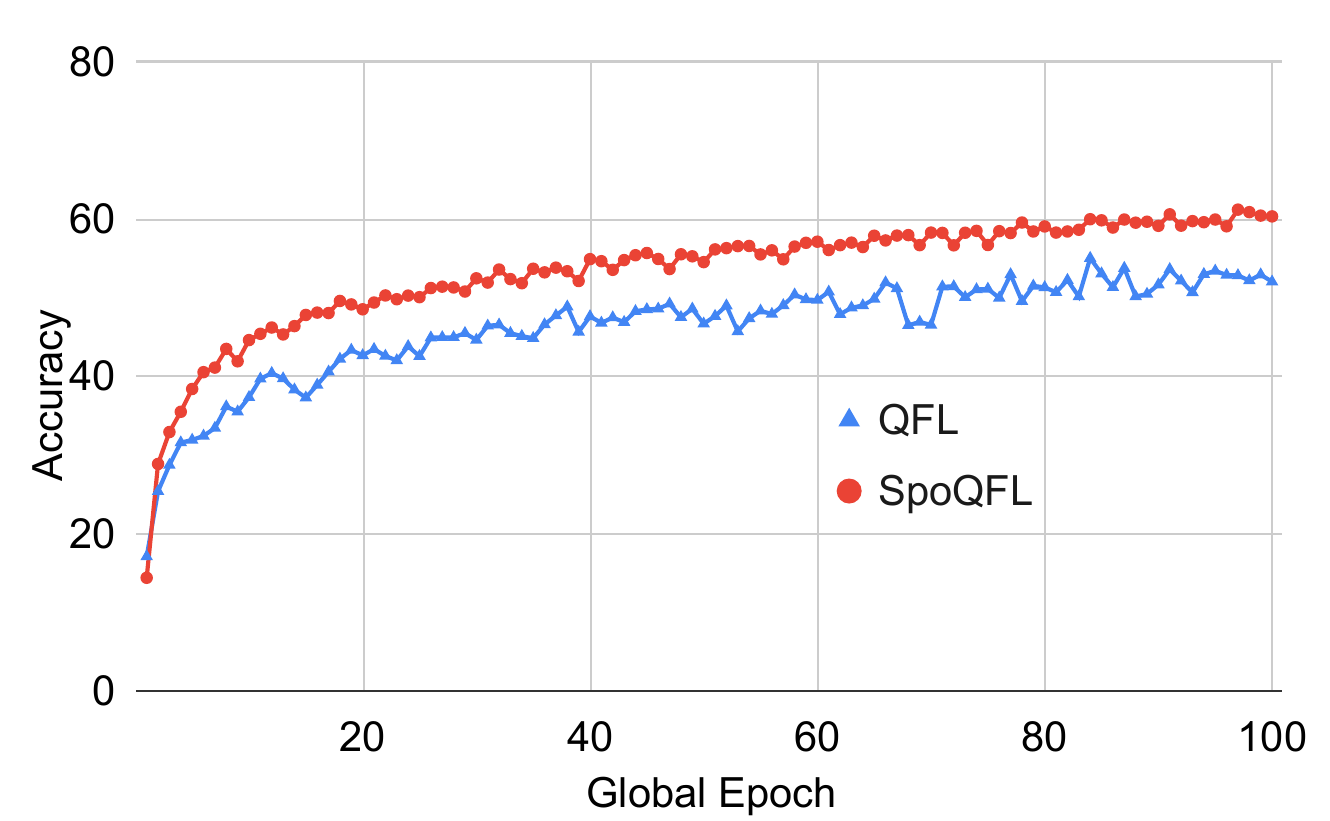}
        \caption{\footnotesize CIFAR-10 Non-IID accuracy.}
    \end{subfigure}
    \caption{Performance comparison between QFL and \textit{SpoQFL} for CIFAR-10 and CIFAR-100 dataset in IID and non-IID data distribution. For adding quantum noise, we set the noise level to $\epsilon = 0.001$.}
    \label{fig: spo}
\end{figure*}

\textbf{Mitigating the noise.} We compare the effectiveness of sporadic learning to mitigate the quantum noise in figure \ref{fig: spo}. The main difference between QFL and \textit{SpoQFL} is how they handle quantum noise heterogeneity in distributed quantum systems.  QFL uses a normal federated learning approach in which model updates are aggregated without accounting for noise changes across quantum devices, resulting in training instability, slower convergence, and reduced accuracy.  \textit{SpoQFL}, on the other hand, uses sporadic learning to effectively mitigate the impact of quantum noise by dynamically changing model updates in response to noise fluctuations.  When compared to QFL, this adaptive mechanism provides more training stability, lower loss, and higher accuracy.  
\begin{table*}[!ht]
    \centering
    \footnotesize
    \begin{tabular}{|c|c|cc|cc|}
    \hline
    \multirow{2}{*}{\textbf{Category}} & \multirow{2}{*}{\textbf{Method}} & \multicolumn{2}{c|}{\textbf{CIFAR-10}} & \multicolumn{2}{c|}{\textbf{CIFAR-100}} \\
    \cline{3-6}
    & & \textbf{Loss Value} & \textbf{Acc.} & \textbf{Loss Value} & \textbf{Acc.} \\
    \hline
    \multirow{4}{*}{Classical FL} 
    & FedAvg \cite{mcmahan2017communication} & 1.2345 & 70.12\% & 2.0456 & 39.45\% \\
    & FedProx \cite{li2020federated} & 1.1876 & 72.34\% & 1.9876 & 40.87\% \\
    & Moon \cite{li2021model} & 1.0984 & 74.56\% & 1.8765 & 42.78\% \\
    & PFL \cite{rahman2024improved} & 1.0567 & 76.89\% & 1.7890 & 45.12\% \\
    \hline
    \multirow{6}{*}{Quantum-Based FL} 
    & QNN \cite{schuld2014quest} & 1.1145 & 73.64\% & 1.8976 & 41.29\% \\
    & QCNN \cite{oh2020tutorial} & 0.7316 & 82.75\% & 1.5053 & 49.43\% \\
    & QFL \cite{chehimi2022quantum} & 0.7440 & 83.67\% & 1.3406 & 51.81\% \\
    & PQFL \cite{shi2024personalized} & 0.7212 & 86.55\% & 1.3760 & 53.81\% \\
    & wpQFL \cite{gurung2024personalized} & 0.6880 & 87.05\% & 1.3690 & 53.94\% \\
    \cline{2-6}
    & \textbf{SpoQFL} & \textbf{0.5748} & \textbf{91.92\%} & \textbf{1.2395} & \textbf{57.60\%} \\
    \hline
    \end{tabular}
    \caption{ Performance comparison between \textit{SpoQFL} and other state-of-the-art approaches.}
    \label{tab:finalcomp}
\end{table*}
\subsection{Comparison with State-of-the-art Approaches} Finally, we compare our proposed \textit{SpoQFL} with other state-of-the-art approaches in Table~\ref{tab:finalcomp}. For comparison, we decide to use a basic FL approach: FedAvg \cite{mcmahan2017communication}, FL for heterogeneous data: FedProx \cite{li2020federated}, model contrastive FL: Moon \cite{li2021model}, personalized FL: PFL \cite{rahman2024improved}, a basic QML structure: QNN \cite{schuld2014quest}, a hybrid QML approach: QCNN \cite{oh2020tutorial}, a basic QFL framework: QFL \cite{chehimi2022quantum}, personalized QFL framework: PQFL \cite{shi2024personalized}, a weighting averaging based personalized QFL: wpQFL \cite{gurung2024personalized}, and our \textit{SpoQFL} algorithm.

%The performance of SpoQFL is compared to both conventional and quantum-based federated learning (FL) approaches in Table~\ref{tab:finalcomp}.  
With an accuracy of 76.89\% on CIFAR-10 and 45.12\% on CIFAR-100, PFL outperforms the other classical FL techniques. Moon comes in second with 74.56\% and 42.78\%, respectively.  FedAvg, the most basic FL approach, falls short, with only 70.12\% accuracy on CIFAR-10 and 39.45\% on CIFAR-100.  The loss values also exhibit a similar pattern, with PFL doing the best among the traditional approaches (1.0567 on CIFAR-10, 1.7890 on CIFAR-100).  However, \textit{SpoQFL} beats all conventional approaches, increasing accuracy over PFL by 15.03\% on CIFAR-10 and 12.48\% on CIFAR-100 while simultaneously lowering loss values, demonstrating that quantum-based FL methods are superior in dealing with complicated datasets even with considering quantum noise.

Among quantum-based FL techniques, wpQFL has the second-best performance, with 87.05\% accuracy on CIFAR-10 and 53.94\% on CIFAR-100, followed by PQFL (86.55\%, 53.81\%) and QFL (83.67\%, 51.81\%).  Classical quantum networks, such as QNN, perform significantly worse, scoring 73.64\% on CIFAR-10 and 41.29\% on CIFAR-100, indicating that hybrid or personalized quantum models are more effective in federated learning.  In terms of loss values, \textit{SpoQFL} reduces the loss from 0.6880 (wpQFL) to 0.5760 on CIFAR-10, a 16.84\% reduction, and from 1.3690 to 1.2395 on CIFAR-100, a 4.15\% improvement.  These results show that \textit{SpoQFL} increases not just accuracy, but also model convergence and stability in a noisy quantum FL environment.

Overall, \textit{SpoQFL} surpasses conventional and quantum-based FL approaches, making it the most effective approach in this comparison.  \textit{SpoQFL} outperforms the strongest classical FL method (PFL) by 15.03\% on CIFAR-10 and 12.48\% on CIFAR-100, as well as the best quantum-based FL method (wpQFL) by 4.87\% and 3.66\%.  Furthermore, the considerable reduction in loss values across datasets demonstrates that \textit{SpoQFL} improves generalization and optimization efficiency, making it the top-performing federated learning approach for CIFAR-based tasks. Therefore, we can conclude that quantum noise plays a significant role in model performance and \textit{SpoQFL} can mitigate the effect of quantum noise, thus improving the overall performance.

\section{Conclusions} 
In this research, we introduced \textit{SpoQFL}, a novel sporadic Quantum Federated Learning (QFL) strategy for addressing quantum noise heterogeneity in distributed quantum systems by dynamically changing model updates in response to noise variations.  Unlike traditional QFL approaches, which suffer from unstable training due to uneven noise across quantum devices, \textit{SpoQFL} enhances convergence stability and learning efficiency, resulting in improved performance.  Our experimental results reveal that \textit{SpoQFL} outperforms previous QFL approaches, reaching up to 4.87\% higher accuracy on CIFAR-10 and 3.66\% on CIFAR-100, while reducing loss by 16.84\% and 4.15\%, respectively.  These findings confirm the effectiveness of sporadic learning in sustaining quantum federated training and leveraging quantum computing benefits. As quantum technology advances, tackling noise-related challenges is essential and \textit{SpoQFL} provides a scalable and effective solution for decentralized quantum learning.  Future research will focus on further enhancements, such as quantum error correction methods for hybrid quantum-classical learning, to improve QFL's resilience for real-world applications.
{
    \small
    \bibliographystyle{ieeenat_fullname}
    \bibliography{main}
}


\end{document}